# Lens–Based Beamformer for Low–Complexity Millimeter–Wave Cellular Systems

M. Ali Babar Abbasi, Vincent F. Fusco, Harsh Tataria, and Michail Matthaiou

*Institute of Electronics, Communications and Information Technology (ECIT),*
*Queen's University Belfast, UK*
*Email: {m.abbasi, h.tataria, v.fusco, m.matthaiou}@qub.ac.uk*

**INTRODUCTION**

Data rate requirements for cellular communications are expected to increase 1000–fold by 2020, compared to 2010. This is mainly because of the rapid increase in the number of wireless devices and data hungry applications per–device. This creates a formidable bandwidth crisis. Milimeter–wave (mmWave) systems with massive multiple–input multiple–output (MIMO) operation are two complementary concepts poised to meet this exploding demand, as verified by the wealth of investigations by both industry and academia. Nevertheless, existing MIMO processing techniques, requiring a dedicated radio–frequency (RF) up/down–conversion chain per antenna, results in prohibitively high complexity and cost of the mmWave prototypes. The primary objective of this paper is to present a complete feasibility study on alternative beamforming techniques targeted to reduce the system complexity without drastically compromising its performance. More precisely, at 28 GHz, we investigate the end–to–end performance of two different lens–based beamforming topologies. Our study is an amalgam of theoretical modelling, electromagnetic design, and prototype manufacturing, yielding a comprehensive mmWave, massive MIMO performance evaluation.

**2–D ROTMAN LENS BASED BEAMFORMER:**

The first proposed beamformer topology is presented in Fig. 1(a). The beamformer comprises of a uniform rectangular array (URA) with two stages of stacked Rotman lenses. URA consists of a coax fed microstrip patch antenna [1] operating at 28 GHz. The antenna has a $|S_{11}| < -10$ dB bandwidth of 3650 MHz and is replicated in 3 row 5 columns to form a URA. Coax feed of each antenna element is connected to high frequency SMP mini connectors [2]. URA is followed by horizontally placed stage–1 lenses and vertically placed stage–2 lenses. All the lenses are developed using tri–focal lens synthesis approach and optimization guidelines presented in [3], [4]. A Detailed description of design parameters is given in [5]. The Stage–1 lenses have 5 array ports and 3 beam ports. The array ports have a one–to–one connection to the 5 antenna elements in each row of the URA. Stage–2 lenses have 3 array ports and 3 beam ports. In summary, the beamformer can support a total number of 9 RF chains, corresponding to beam ports of 3 × stage–2 lenses. The beamformer is capable of steering in both azimuthal and elevation planes depending upon the excitation of a particular RF chain, as show in Fig. 1(b). Vertically placed lenses are responsible of creating a phase ramp in elevation plane while the horizontally placed lenses do the same in azimuth plane. We designed and simulated the entire geometry, presented in Fig. 1(a), in three sequential steps using three numerical simulation software packages namely: Keysight advanced design system (ADS) [6], computer simulation technology microwave studio (CST MS) and CST design studio (CST DS) [7].

*Step – 1*: The URA was designed and simulated in CST MS and the performance of all the radiating elements were recorded. The URA unit cell was than optimized for high efficiency operation (88%) at 28 GHz in the presence of other antenna elements and surface mount SMP connectors. Similarly, Rotman lens for both stages was designed in separate CST MS files.
*Step – 2*: The Lens structures were than separately imported in ADS and phased aligned 50 Ohm transmissions were designed to connect the Lens's body to the edge of the PCB, such that a precise connector placement is ensured. All the transmission lines were designed keeping in mind the physical dimensions required to realize the stacked configuration of Fig. 1(a). Next both Rotman lenses with the transmission lines were imported as separate files of CST MS. Simulations were carried out using finite–difference–time–domain (FDTD) method, while including the edge mount SMP–M connectors with the Rotman lenses.
*Step – 3*: Finally, three separate CST MS studio files, i.e. for URA, stage–1 and stage–2 Rotman lenses were combined as an assembly in CST DS. Using co–simulation setup, we evaluated the far–field beamforming capabilities of the proposed beamformer topology. The final results are shown in Fig. 1(b) and summarized in Table 1.

Note that the Rotman lens have a capability of replicating the phase–shifter network, however, due to inherent defects in the lens body, phase–shifting is not always accurate. These defects are not covered in this study and further details can be found in our concurrent work [8], [9]. To ensure equal power beams in azimuth and elevation zones, we carried out a power equalization in RF chain excitations. It can be noticed from Table 1 that the half power beam width (HPBW) along azimuth is lower than the elevation because of unequal antenna elements in URA rows and columns. In azimuth plane, as the beam tilts from the broad side direction, the HPBW increases by a factor of ~5°. The opposite is true for the elevation plane where HPBW decrease of ~4° is seen, as we move away from the broadside direction. We noticed that this imbalance between the HPBW is a result of power loses that are different for every individual lens port and transmission line. We also noticed that the overall performance of stage–2 lens is better compared to stage–1 lens. Once reason for that is the physical separation between beam ports in stage–2 lens is wider, providing a better port isolation. Note that on average, a 0 dB signal at the input beam ports faces an overall loss of ~8.52 dB before being able to radiate from the URA.

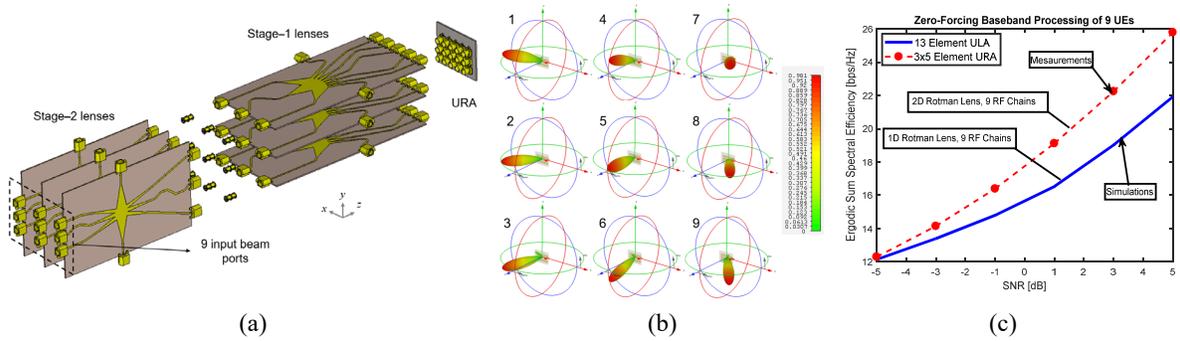

Fig. 1. Two–stage stacked Rotman lens and URA based beamformer for azimuth and elevation zone coverage. (b) Far–field realized gain in linear scale. (c) Ergodic sum spectral efficiency estimation comparison of the proposed topology with single–staged Rotman lens based beamformer.

Table 1. Simulated results of two–stage stacked Rotman lens beamformer when RF chains are excited

| RF chain | 1 | 2 | 3 | 4 | 5 | 6 | 7 | 8 | 9 |
|---|---|---|---|---|---|---|---|---|---|
| $P_{in}$ (Watt) | 1.89 | 1.00 | 1.91 | 2.63 | 1.60 | 2.67 | 1.97 | 1.05 | 1.99 |
| HPBW along azimuth | 30.6° | 28.6° | 29.7° | 24.2° | 24.6° | 24.5° | 30.3° | 28.1° | 29.8° |
| HPBW along elevation | 42.4° | 47.0° | 44.7° | 29.0° | 35.4° | 31.2° | 42.5° | 47.1° | 44.3° |

Using the beamformer and far–field pattern measurements, we estimated the end–to–end spectral efficiency for an uplink massive MU–MIMO system. A Detailed description on the system model and propagation channel simulation setup is provided in [5]. The results presented in Fig. 1(c) demonstrates the superiority of a two–stage stacked Rotman lens topology compared to single stage Rotman lens based beamformer used at MU-MIMO cellular base station. Contrary to single–stage Rotman lens based beamformer [10]–[14], the proposed topology is capable of steering in both azimuth of elevation zones, resulting in better separate–ability of dominant multi–path components. The difference in ergodic sum spectral efficiency is low for lower SNR values, but it increases with an increase in SNR, due to the second dimension by the proposed topology.

**HIGH DIRECTIVITY DIELECTRIC LENS BEAMFORMER:**

The second type of beamformer proposed in this study is the constant-$\varepsilon_r$ lens beamformer. The lens structure is based on a homogenous material with a constant permittivity, permeability, conductivity, and mass density. The working principles of such lens type had thoroughly been investigated in the past [15]–[17]. In the paper, we present an extension to this technique considering the mmWave MIMO array requirements. Assume an EM plane wave exciting an antenna aperture, similar to a MIMO URA receiver. Spherical dialectics have a capacity to focus this energy to a single point, however, this energy convergence cannot be identical for an entire plane wave. Consider, for example, an EM energy path represented by rays *a, b* and *c* hitting the lens at multiple points, as shown in Fig. 2(a). The total *path – length* for ray *a* is more than that for ray *b*, and ray *c*. In ideal conditions, focusing of these rays to a single point at the same time instance is not possible. Consider now a careful decrease in the EM wave's phase velocity for the ray *c*, and timing it to match with the total phase velocity of ray *a*. This way, a plane wave can be converted to a focal point using a spherical lens with carefully selected refractive index. Conversely, placement of a radiating element at this focal point will result in a plane EM wave leaving the lens. For different angle-of-arrivals (AOA), the EM wave focusing is not possible on a plane, as in URA, so Petzval curvature needs to be considered as a focusing contour [18]. Constant-$\varepsilon_r$ lens provides a unique flexibility of radiating element placement outside the lens surface, making it a low-complexity, low-cost and a scalable solution. It is important to mention that almost all the hardware anomalies associated to an optical lens hold valid with this lens type (such as astigmatism and coma–aberration etc.). However, due to the technological advancement of modern microwave engineering, a number of anomalies are intelligible and resolvable.

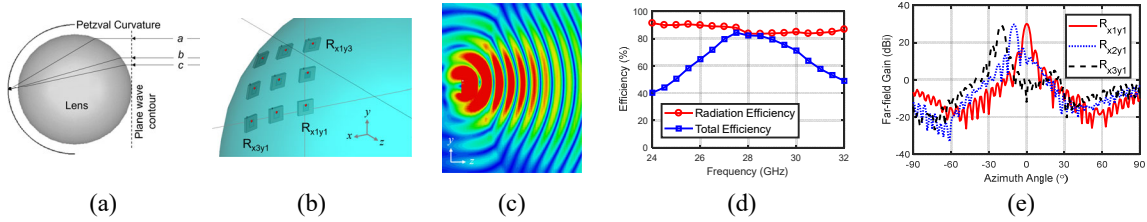

(a) (b) (c) (d) (e)

Fig. 2. Ray tracing representation of s projected by a plane wave, (b) geodesic placement of radiating elements, (c) Normalized E–field plot when a single radiating element (a patch antenna) is excited. (d) Constant-$\varepsilon_r$ lens antenna efficiency when element $R_{x1y1}$ is excited. (e) Far–field gain corresponding to elements excitation along the $xz$–plane.

We designed a lens using dielectric material with $\varepsilon_r$ = 2.53. First, we evaluated the optimum Petzval curvature in CST MS [7] using Multilevel Fast Multipole Method (MLFMM) solver. We designed and placed a patch antenna as a radiator at the focal point. The assembly is capable of beamforming in single direction, which is not enough for MIMO operation. We than evaluated the electrical field minima on the Petzval curvature, and placed multiple antennas to realize multi-beams using the same lens aperture. Electric field minima were considered to ensure the least possible mutual coupling between neighbouring radiating elements. We simulated the entre assembly in CST MS using FDTD method, the results are present in Fig. 2(c) – (e). It can be noticed from Fig. 3(c) that when an individual patch antenna radiator is excited, EM energy enters the lens surface following the condition $\lambda_{lens} < \lambda_{air}$. In addition, there are surface waves outside lens that *creep* along the lens surface, which can create an RF cross-talk with other radiating elements. To avoid this, highly directive and isolated radiators can be used, such as wave guides and horn antennas. However, we observed that when such radiators are used to excite the lens, physical location sensitivity increases to an extent that a ~100 μm displacement of the radiator along $z$–axis can cost a loss of ~1.5 dB in the peak far-field gain. Thus, a high precision hardware manufacturing is required.

We tailored the beamformer in Fig. 2(b) for 28 GHz operation. Highest radiation efficiency of the assembly was achieved when the lens radius is 66 mm, focal radius is 70.5 mm and inter element geodesic spacing along $xz$– and $yz$– planes is ~13 mm. With an equal power excitation, the HPBW of the beamformer when $R_{x1y1}$, $R_{x2y1}$ and $R_{x3y1}$ are excited is 4.0°, 4.2° and 4.3° (Fig. 2(e)). The HPBW corresponding to any of the elements shown in Fig. 2(b) stayed within the range 4.0° to 4.5°, with a gain value deviation of ±0.4 dB. Very low HPBW and a very high gain makes the proposed architecture a very lucrative choice for high directivity demands of mmWave MIMO RF front-end.

**CONCLUSION AND FUTURE PROSPECTIVES:**

Two beamformer topologies for the RF front-end of mmWave MIMO hybrid architecture are presented. Both topologies are studied and optimized for a successful operation at 28 GHz. The first topology comprises of a URA with a two–stages of stacked Rotman lenses. The second topology is based on a constant-$\varepsilon_r$ lens excited by patch antenna radiators placed at the Petzval curvature. In the next step, we are planning to estimate an accurate theoretical upper limit of both beamforming topologies for MU-MIMO scenarios. Practical demonstration of both systems in combination with baseband SP is one of the future goals.